\documentclass[conference]{IEEEtran}
\IEEEoverridecommandlockouts 
\usepackage{siunitx}
\usepackage{cite}
\usepackage[hidelinks]{hyperref}
\usepackage{amsmath,amssymb,amsfonts}
\usepackage{graphicx}
\usepackage{textcomp}
\usepackage{xcolor}
\usepackage{booktabs}
\usepackage{multirow}
\usepackage{url}
\usepackage[section]{placeins}
\begin{document}
\title{\vspace{0.2in}\fontsize{19.2}{22.5}\selectfont
Risk-Sensitive Specialist Routing for Volatility Forecasting}

\author{
\vspace{-1em}
Tenghan Zhong$^{1}$\thanks{$^{1}$Tenghan Zhong is with the Department of Mathematics, University of Southern California, Los Angeles, CA, USA (e-mail: tenghanz@usc.edu).}
}

\maketitle

\begin{abstract}
Volatility forecasting becomes challenging when market conditions shift and model performance varies across market states. Motivated by this instability, we develop a risk-sensitive specialist routing framework for ETF volatility forecasting. The framework uses online risk-sensitive evaluation and state-dependent gating to combine different forecasting specialists across calm and stressed market states. Using a daily panel of six ETFs under a rolling walk-forward design, we find that the strongest forecaster is regime-dependent rather than stable across all states. Relative to the rolling-best baseline, the proposed routing framework reduces high-volatility forecast loss by about 24\% and underprediction loss by about 22\%. These results suggest that specialist routing provides a practical forecasting architecture that adapts to changing market conditions.
\end{abstract}


\section{Introduction}

Daily volatility forecasting plays a significant role in financial decision support, with applications in risk management, portfolio allocation, and derivative pricing. In the literature, this problem has long been studied through conditional heteroskedasticity models \cite{engle1982arch} and realized-volatility methods \cite{andersen2003modeling}, and more recently through machine-learning approaches \cite{gunnarsson2024review}. Despite this broad modeling effort, forecasting performance in practice often becomes unstable as market conditions evolve, and the relative performance of models can vary substantially across volatility regimes \cite{marcucci2005}. This challenge is especially relevant for Exchange-Traded Fund (ETF) volatility forecasting \cite{zhu2019fxi}, where predictive performance depends not only on model design and information sets, but also on the volatility state \cite{maki2024downside}.

We therefore study next-day ETF volatility forecasting through risk-sensitive specialist routing rather than single-model dominance. The goal is not only to improve average accuracy, but also to adapt model selection and forecast combination as model performance changes across market states. Motivated by evidence that regime-dependent dynamics matter for volatility forecasting \cite{ding2025regimeswitching}, we develop an online framework that evaluates models with a risk-sensitive criterion and adaptively routes them over time.

This paper makes three main contributions. First, it develops an online forecasting architecture that scores candidate models in a risk-sensitive way, routes specialists across market states, and combines them into a final forecast. Second, using a rolling walk-forward design on a multi-ETF daily panel, it shows that the strongest forecaster varies systematically across regimes. Third, it shows that the proposed framework improves stressed-regime performance and remains competitive overall relative to benchmark forecasts.

\section{Related Work}

Among traditional benchmarks, autoregressive conditional heteroskedasticity (ARCH)-type models remain central in volatility forecasting, while fractionally integrated GARCH (FIGARCH) \cite{baillie1996figarch} and heterogeneous autoregressive realized volatility (HAR-RV) \cite{corsi2009har} continue to play important empirical roles.

More recent work has extended this literature toward machine learning and deep learning, including tree-based methods \cite{christensen2023ml}, panel-data machine learning \cite{zhu2023panel}, and deep architectures built from high-frequency inputs \cite{moreno2024deepvol}. However, their gains are not uniform across assets, forecast horizons, and market conditions \cite{branco2024anything}.

Related lines of work also study how forecasting systems should adapt to changing market states and select among competing predictors. In econometrics, Bayesian and dynamic model averaging provide probabilistic frameworks for combining forecasts over time \cite{liu2009bma,raftery2010dynamic}, while Markov-switching realized-volatility models incorporate latent market states directly into volatility prediction \cite{wang2019mshar}. Other work considers more explicit model-selection rules for volatility forecasting, including false-discovery-rate-based approaches \cite{hassanniakalager2024fdr}. From a broader machine-learning perspective, mixture-of-experts models provide a general mechanism for assigning specialized predictors to different input regions \cite{jacobs1991adaptive}, while gating and hierarchical expert architectures extend this idea to state-dependent routing structures \cite{jordan1994hierarchical}.

Our framework differs from this literature in three respects. First, unlike dynamic model averaging, it scores each model by its loss gap relative to the current best performer and reweights recent history by regime similarity. Second, unlike mixture-of-experts architectures with learned gates, it uses pre-specified specialist pools and observable market-state variables, which keeps the routing rule interpretable in a limited-sample walk-forward setting. Third, its scoring rule is explicitly risk-sensitive: it penalizes relative underprediction, so routing decisions reflect not only overall fit but also protection against underforecasting.

\section{Framework and Method}

\begin{figure}[htp]
\centering
\includegraphics[width=1.00\columnwidth]{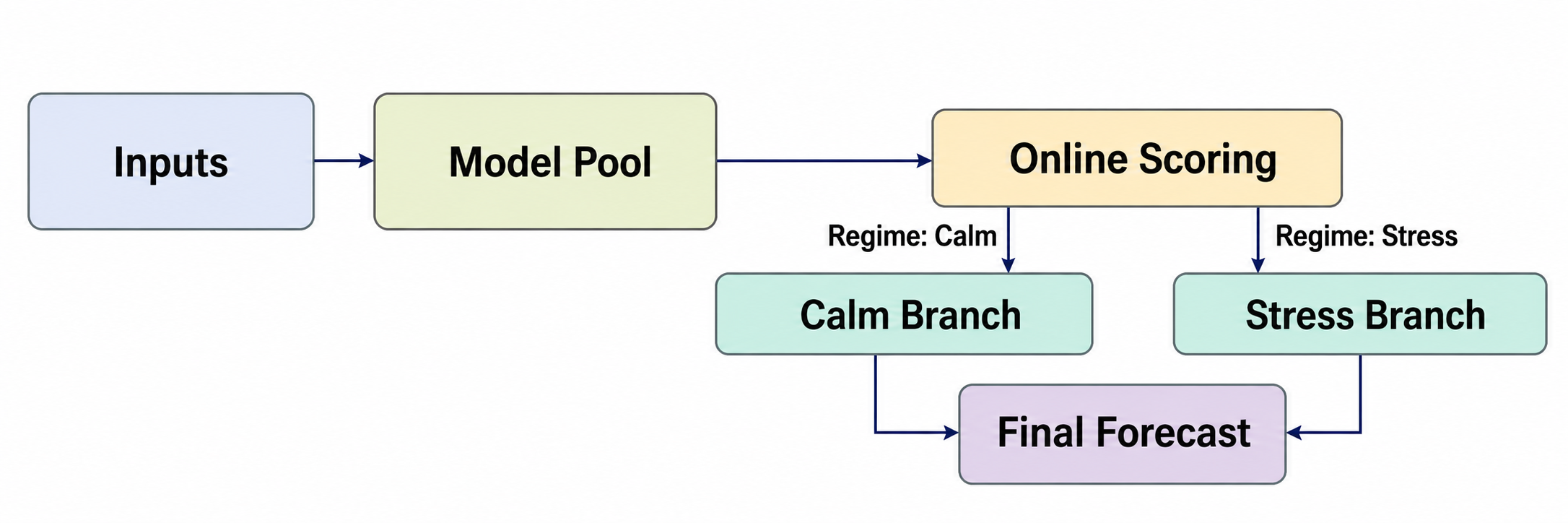}
\caption{Risk-sensitive specialist routing architecture.}
\label{fig:framework}
\end{figure}

The forecasting framework has four main stages: input and model-pool setup, online scoring, routing into calm and stress specialist branches, and final forecast combination. Fig.~\ref{fig:framework} summarizes the workflow.

\subsection{Inputs, Target, and Model Pool}

The first stage specifies the information set for the next-day forecast.

The forecasting target is next-day realized variance, proxied by the squared Garman--Klass volatility proxy \cite{garman1980estimation}. Let $\mathrm{gk\_proxy}_{t+1}$ denote the next-day Garman--Klass volatility proxy, and define
\[
y_{t+1} = \bigl(\mathrm{gk\_proxy}_{t+1}\bigr)^2.
\]

At each forecast date, forecasts are generated from a model pool that includes both econometric and machine-learning forecasters. In the baseline implementation, this pool consists of HAR-RV, GARCH(1,1)-t, FIGARCH(1,1)-t, GRU (gated recurrent unit), and XGBoost (extreme gradient boosting).

\subsection{Online Specialist Scoring}

The second stage of the framework evaluates candidate models according to recent forecasting performance. Let $\hat y_t^{(m)}$ denote the forecast for $y_{t+1}$ issued by model $m$ at time $t$. We use a risk-sensitive loss that combines quasi-likelihood (QLIKE) loss \cite{patton2011volatility} with an additional underprediction penalty.

For a candidate forecast $\hat y_t^{(m)}$ from model $m$, define the underprediction term by
\[
U_{t+1}^{(m)}
=
\left(
\frac{\max\!\left(y_{t+1}-\hat y_t^{(m)},0\right)}{y_{t+1}}
\right)^2.
\]
This term is positive only when realized variance exceeds the forecast and penalizes large relative underforecasting. We refer to $U_{t+1}^{(m)}$ as the underprediction loss below.

We then define the total loss as
\begin{equation*}
L_{t+1}^{(m)}
=
\mathrm{QLIKE}\!\left(y_{t+1}, \hat y_t^{(m)}\right)
+
\lambda_{\mathrm{under}}\,U_{t+1}^{(m)}.
\end{equation*}
The QLIKE loss is defined by
\[
\mathrm{QLIKE}(y,\hat y)=\frac{y}{\hat y}-\log\!\left(\frac{y}{\hat y}\right)-1,
\]
and $\lambda_{\mathrm{under}} > 0$ controls the weight placed on underprediction.

Next, we compare each model with the best active model on the same date by forming an excess loss. Specifically,
\[
R_{t+1}^{(m)}
=
L_{t+1}^{(m)}
-
\min_{j \in \mathcal{A}_t} L_{t+1}^{(j)},
\]
where $\mathcal{A}_t$ denotes the active candidate pool. Thus, the quantity $R_{t+1}^{(m)}$ measures the excess loss of model $m$ relative to the best active model, with smaller values indicating better relative performance.

To obtain an online score, we aggregate excess losses over the most recent 252 evaluation dates, denoted by $\mathcal{H}_t$. The score of model $m$ at time $t$ is
\[
S_t^{(m)}
=
\frac{\sum_{s \in \mathcal{H}_t} w_t(s) R_{s+1}^{(m)}}
{\sum_{s \in \mathcal{H}_t} w_t(s)},
\]
where the weights $w_t(s)$ place more emphasis on recent observations and on past dates with market states similar to the current one. Specifically,
\[
w_t(s)\propto
\exp\!\left(-\gamma_{\mathrm{time}}(t-s)\right)
\exp\!\left(-\frac{\|z_t-z_s\|^2}{\gamma_{\mathrm{reg}}^2}\right),
\]
where $z_t$ standardizes prior-history values of
$\log\mathrm{VIX}$, $\log(\mathrm{VIX}/\mathrm{VXV})$, 20-day RV,
20-day volatility-of-volatility, the term spread, and the HY spread. The parameters
$\gamma_{\mathrm{time}}$ and $\gamma_{\mathrm{reg}}$ control time decay
and regime similarity. Lower $S_t^{(m)}$ indicates better recent
performance in similar states.

\subsection{State-Dependent Routing Set}

The third stage uses the online scores to determine which models remain competitive under the current market state. Rather than selecting a single best model, we form a routing set and use it to construct calm and stress specialist branches.

To balance local adaptivity and stability, we use a threshold that shrinks a regime-local quantile toward a global quantile:
\[
\tau_t
=
(1-\eta_t)\tau_t^{\mathrm{global}}
+
\eta_t \tau_t^{\mathrm{local}},
\]
where $\tau_t^{\mathrm{global}}$ is the $(1-\alpha)$-quantile (with $\alpha=0.10$) of recent routing scores, $\tau_t^{\mathrm{local}}$ is the corresponding regime-local weighted quantile, and $\eta_t \in [0,0.80]$ increases with the effective local sample size. Thus, $\tau_t$ retains models whose current scores remain close to recent competitive performance.

The routing set is then defined as
\[
\mathcal{M}_t
=
\left\{
m \in \mathcal{A}_t : S_t^{(m)} \le \tau_t
\right\},
\]
with fallback to the top-ranked available model if the set is empty. In the default design, at most one model is retained in calmer conditions and at most two models are retained in more stressed conditions, so the routing set is narrower in calmer states and slightly wider under stress.

\subsection{Final Forecast Construction}
The fourth stage converts the routing set into branch-level forecasts and then combines these branches into the final forecast. We begin with two specialist pools for calmer and more stressed market conditions:
\begin{equation*}
\begin{aligned}
\mathcal{P}_{\mathrm{calm}}
&=
\{\mathrm{GRU}, \mathrm{HAR\mbox{-}RV}, \mathrm{XGBoost}\}, \\
\mathcal{P}_{\mathrm{stress}}
&=
\{\mathrm{GARCH\mbox{-}t}, \mathrm{FIGARCH}, \mathrm{HAR\mbox{-}RV}\}.
\end{aligned}
\end{equation*}

The calm pool emphasizes nonlinear and realized-volatility models, while the stress pool emphasizes Student-$t$ and long-memory volatility models; HAR-RV anchors both pools. The pools were pre-specified and not modified after evaluation. At each forecast date, each branch draws from the models that belong to both its specialist pool and the routing set $\mathcal{M}_t$. Thus, the calm branch is based on $\mathcal{P}_{\mathrm{calm}} \cap \mathcal{M}_t$, and the stress branch is based on $\mathcal{P}_{\mathrm{stress}} \cap \mathcal{M}_t$. If either intersection is empty, the corresponding branch falls back to the top-ranked available model(s) within its specialist pool. The calm branch uses the median of the retained forecasts, whereas the stress branch uses the 75th percentile after winsorizing them at their empirical 10th and 90th percentiles. This yields $\hat y_t^{\mathrm{calm}}$ and $\hat y_t^{\mathrm{stress}}$.

We then compute a stress score $p_t \in [0,1]$ from the standardized market-state vector $z_t$ by applying a logistic transform to a signed linear index of the state variables:
\[
p_t = \sigma\!\left(\frac{\sum_k \alpha_k z_{t,k}}{\sqrt{\sum_k \alpha_k^2}} - c_0 \right),
\]
where $z_t$ is as above, $\alpha=(1,0.75,0.85,0.85,-0.50,0.60)$ follows that variable order, $c_0=0.20$, and $\sigma(\cdot)$ is the logistic function. Larger values of $p_t$ place more weight on the stress branch, so that
\[
\hat y_t^{\mathrm{combo}}
=
(1-p_t)\hat y_t^{\mathrm{calm}}
+
p_t \hat y_t^{\mathrm{stress}}.
\]

To stabilize the final forecast across market conditions, we next form separate low-state and high-state combinations using the rolling-best benchmark and the specialist branches. Let $\hat y_t^{\mathrm{roll}}$ denote the rolling-best benchmark forecast. We then define a low-state branch and a high-state branch:
\[
\hat y_t^{\mathrm{low}}
=
(1-\rho)\hat y_t^{\mathrm{roll}}
+
\rho \hat y_t^{\mathrm{calm}},
\]
\[
\hat y_t^{\mathrm{high}}
=
(1-\kappa)\hat y_t^{\mathrm{combo}}
+
\kappa \hat y_t^{\mathrm{stress}},
\]
where $\rho,\kappa \in [0,1]$ control the amount of branch blending.

The final forecast is obtained by using a second gate to blend the low-state and high-state branches:
\[
\omega_t
=
\sigma\!\left(\frac{p_t-c}{b}\right),
\qquad
\hat y_t
=
(1-\omega_t)\hat y_t^{\mathrm{low}}
+
\omega_t \hat y_t^{\mathrm{high}},
\]
where $\omega_t \in [0,1]$ increases with the stress score, $c$ controls the gate midpoint, and $b$ controls the smoothness of the transition between the two branches.

Finally, letting $\hat y_t^{\mathrm{HAR}}$ denote the HAR-RV forecast and letting $D_t$ denote forecast disagreement among the routed models, measured as the interquartile range divided by the absolute median with an $\varepsilon$ safeguard, we apply a conditional HAR floor, where $p_{\mathrm{floor}}$ and $d_{\mathrm{floor}}$ are fixed trigger thresholds for $p_t$ and $D_t$, respectively:
\[
\hat y_t
\leftarrow
\begin{cases}
\max\!\left\{\hat y_t,\hat y_t^{\mathrm{HAR}}\right\},
& \text{if } p_t \ge p_{\mathrm{floor}} \text{ or } D_t \ge d_{\mathrm{floor}}, \\[0.4em]
\hat y_t,
& \text{otherwise}.
\end{cases}
\]
In the baseline implementation, the HAR floor is activated when $p_t \ge 0.65$ or $D_t \ge 0.20$.

\section{Experimental Design}

\subsection{Data and Inputs}

We use two daily data sources: a multi-asset ETF market panel and a macro-financial dataset. The ETF panel provides the return and volatility information used to construct the next-day realized variance target and ETF-based forecasting variables, while the macro file provides market-state variables. The ETF-based variables include lagged realized variance, return-based variables, and realized-volatility features; the macro variables include transformed VIX information, yield-curve quantities, and credit-spread information. All series are aligned by trading date and sorted chronologically.

The analysis considers six ETFs spanning U.S.\ equities (SPY, QQQ, IWM), emerging markets (EEM), gold (GLD), and long-term Treasuries (TLT). The market panel covers February 2015 through December 2025. After the initial 504-day history-accumulation period, the common evaluation sample runs from March 2017 to December 2025, yielding 2,219 trading days per asset. Each ETF is evaluated separately under the same forecasting pipeline before cross-asset aggregation.

\subsection{Walk-Forward Protocol}

Forecasting begins only after the minimum history requirement is met and the next-day target and regime variables are available. In the main specification, both the minimum training history and the rolling training window are set to 504 trading days. HAR-RV, GRU, and XGBoost are re-estimated every 21 trading days on the rolling window, whereas GARCH-t and FIGARCH are re-estimated at each forecast date on the same window. Forecasts are then issued sequentially; after model forecasts are available, routing only updates recent scores, a threshold, and scalar gates.

The benchmark and routing components use separate historical windows. The rolling-best benchmark is defined by the model with the lowest recent average risk-sensitive loss over a 252-day comparison window. The static-best benchmark is defined by the model with the lowest average risk-sensitive loss over an initial 252-day comparison period and is then held fixed throughout the remaining evaluation sample. These adaptive baselines do not introduce separate retraining schemes; they select among candidate forecast streams generated under the same walk-forward protocol. At each forecast date, the routing threshold uses the most recent 252 dates; fixed settings are $\lambda_{\mathrm{under}}=1$, a 63-day half-life, $\gamma_{\mathrm{reg}}=1.5$, $\rho=0.25$, $\kappa=0.35$, $c=0.55$, $b=0.12$, GRU (20-step, 16 hidden units), and XGBoost (250 trees, depth 3, learning rate 0.05).

To summarize performance across different market conditions, we group the evaluation sample into low-, mid-, and high-volatility periods using the Garman--Klass realized-volatility proxy. This grouping is used only for reporting. The online routing rule itself uses the market-state variables and stress score defined in the method section.

\subsection{Methods and Baselines Compared}

We compare the proposed routing architecture with both single-model and adaptive baselines. The single-model comparators are HAR-RV, GARCH-t, FIGARCH, GRU, and XGBoost. The adaptive baselines are static-best, rolling-best, and a naive VIX-switch rule that uses GARCH-t when raw VIX exceeds 20 and GRU otherwise. For the proposed architecture, the final forecast is reported, and the internal routing behavior is summarized through the routing diagnostics.

\subsection{Evaluation Metrics and Cross-Asset Summaries}

Performance is evaluated primarily on the next-day realized variance scale using QLIKE, which matches the framework's target variable. To avoid undefined or numerically explosive QLIKE values, all forecast variances are truncated below at a common small positive constant before losses are computed. Since the framework is risk-sensitive, we also report the analogous relative underprediction loss on the volatility scale.

For the routing layer, we report four diagnostics, defined in Table~\ref{tab:routing_metric_defs}: calm- and stress-branch usage rates, selected regret, and miss-best rate. In the cross-asset analysis, results are summarized by the median across the six ETFs, both overall and within each volatility regime. Methods are compared using QLIKE and underprediction loss.

\begin{table}[htp]
\centering
\caption{Routing diagnostics used in the evaluation}
\label{tab:routing_metric_defs}
\scriptsize
\setlength{\tabcolsep}{4pt}
\begin{tabular}{p{2.7cm}p{5.2cm}}
\toprule
Metric & Definition \\
\midrule
Calm-/stress-branch usage rate
& Share of dates on which the corresponding branch uses at least one routed model. \\

Selected regret
& Loss gap relative to the best-performing active model. \\

Miss-best rate
& Share of dates on which the routing set misses the best model. \\
\bottomrule
\end{tabular}
\end{table} 

To assess whether pairwise loss differences are statistically significant at the asset level, we additionally conduct Diebold--Mariano tests of predictive accuracy \cite{diebold1995comparing}, using a Newey--West heteroskedasticity- and autocorrelation-consistent (HAC) variance estimate for the loss differential \cite{newey1987hac}.

\section{Results}

We begin by examining whether the strongest forecaster is stable across market states or instead varies systematically with the volatility regime.

\subsection{Regime Dependence of the Best Forecaster}

\begin{figure}[htp]
\centering
\includegraphics[width=0.9\columnwidth,trim=0 8 0 80,clip]{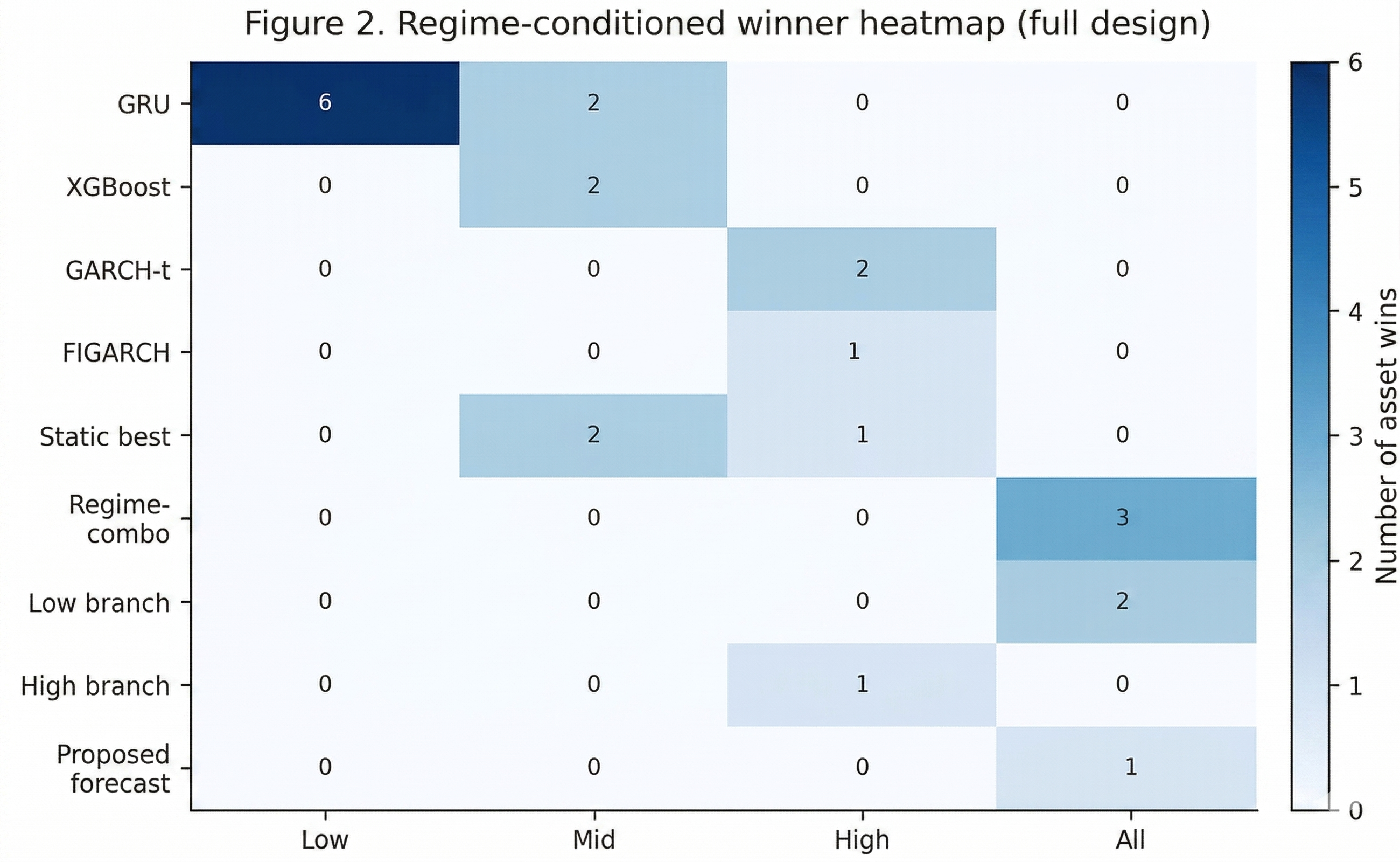}
\caption{Regime-conditioned winner heatmap.}
\label{fig:winner_heatmap}
\end{figure}

Fig.~\ref{fig:winner_heatmap} shows that the strongest forecaster is regime-dependent. For each asset and regime, we identify the model that most often attains the lowest loss. The heatmap then reports, for each regime, how many of the six assets are won by each model. Here, \emph{All} denotes the full evaluation sample. In the low-volatility regime, GRU is the winning model for all six assets. In the high-volatility regime, the strongest methods shift toward stress-oriented econometric specialists, especially GARCH-t and FIGARCH, while GRU deteriorates sharply. Consistent with regime-switching volatility evidence, this pattern indicates that no single model class is uniformly strongest across regimes.

Table~\ref{tab:main_results} confirms the same pattern among the single-model forecasters. GRU has the lowest median QLIKE in the low and mid regimes, whereas GARCH-t and FIGARCH become the strongest single models in the high-volatility regime.

\begin{table}[htp]
\centering
\caption{Cross-asset median forecast losses by regime.}
\label{tab:main_results}
\footnotesize
\setlength{\tabcolsep}{4pt}

\begin{tabular}{l
                S[table-format=1.4]
                S[table-format=1.4]
                S[table-format=1.4]
                S[table-format=1.4]}
\toprule
Method & {Overall} & {Low} & {Mid} & {High} \\
\midrule

\multicolumn{5}{l}{\textit{Panel A: QLIKE}} \\
\addlinespace[0.2em]
Proposed forecast & 0.3232 & 0.4772 & 0.1770 & 0.3685 \\
Rolling-best      & 0.3653 & 0.4516 & 0.1624 & 0.4834 \\
Static-best       & 0.4424 & 0.3758 & 0.1442 & 0.6694 \\
VIX-switch        & 0.4013 & 0.3090 & 0.2863 & 0.6221 \\
HAR-RV            & 0.3603 & 0.4451 & 0.1377 & 0.4572 \\
GRU               & 0.4587 & 0.2362 & 0.1259 & 1.0013 \\
GARCH-t           & 0.5689 & 0.9455 & 0.4574 & 0.2879 \\
FIGARCH           & 0.5775 & 0.9455 & 0.4784 & 0.2958 \\
XGBoost           & 0.5431 & 0.6313 & 0.1451 & 0.8356 \\

\midrule
\addlinespace[0.5em]
\multicolumn{5}{l}{\textit{Panel B: Underprediction loss}} \\
\addlinespace[0.2em]
Proposed forecast & 0.0192 & 0.0000 & 0.0039 & 0.0474 \\
Rolling-best      & 0.0245 & 0.0003 & 0.0068 & 0.0609 \\
Static-best       & 0.0262 & 0.0001 & 0.0071 & 0.0675 \\
VIX-switch        & 0.0297 & 0.0018 & 0.0175 & 0.0650 \\
HAR-RV            & 0.0238 & 0.0001 & 0.0063 & 0.0643 \\
GRU               & 0.0455 & 0.0018 & 0.0201 & 0.1103 \\
GARCH-t           & 0.0183 & 0.0000 & 0.0119 & 0.0327 \\
FIGARCH           & 0.0184 & 0.0000 & 0.0124 & 0.0328 \\
XGBoost           & 0.0306 & 0.0000 & 0.0021 & 0.0891 \\
\bottomrule
\end{tabular}

\vspace{0.2em}
\parbox{\columnwidth}{\footnotesize
\textit{Notes:} Lower values are better for both metrics.
}
\end{table}

The proposed forecast achieves the lowest overall median QLIKE, but this advantage is concentrated in the high-volatility regime. GRU is strongest in the low and mid regimes, whereas GARCH-t and FIGARCH are strongest in the high regime. Relative to the rolling-best benchmark, however, the proposed forecast reduces QLIKE by about 24\% and underprediction loss by about 22\% in the high-volatility regime.

We also compare the proposed forecast with a naive VIX-switch baseline. Although this simple rule is competitive in the low-volatility regime, it remains materially worse overall and in the high-volatility regime in both QLIKE and underprediction loss.

\subsection{Routing-Set Diagnostics}

To clarify how the routing layer behaves across regimes, Table~\ref{tab:routing_diag} reports routing diagnostics by regime. Two patterns stand out. First, stress-branch usage rises from 0.577 in the low-volatility regime to 0.836 in the high-volatility regime, consistent with the intended stressed-state routing behavior, while calm-branch usage remains non-negligible because the routing layer still retains at least one calm-pool model on many dates. Second, miss-best rates are relatively high, whereas selected regret remains much smaller. In this setting, selected regret is therefore the more informative routing diagnostic. This suggests that the routing layer adds value mainly by avoiding clearly unfavorable selections and retaining a small set of competitive models, rather than by identifying the exact best-performing model on each date.

\begin{table}[htp]
\centering
\caption{Routing diagnostics by volatility regime.}
\label{tab:routing_diag}
\scriptsize
\setlength{\tabcolsep}{4pt}
\begin{tabular*}{\columnwidth}{@{\extracolsep{\fill}}lcccccc@{}}
\toprule
Regime & Calm usage & Stress usage & Selected regret & Miss-best rate \\
\midrule
Overall & 0.825 & 0.714 & 0.208 & 0.671 \\
Low     & 0.771 & 0.577 & 0.170 & 0.706 \\
Mid     & 0.870 & 0.730 & 0.110 & 0.672 \\
High    & 0.833 & 0.836 & 0.331 & 0.639 \\
\bottomrule
\end{tabular*}
\end{table}

\subsection{Ablation and Robustness Checks}

Table~\ref{tab:focused_ablation} reports three focused robustness ablations: removing the risk-sensitive underprediction term, removing the additional stress-branch weight in the high-state branch, and removing the conditional HAR floor. We report overall, high-regime, and tail underprediction losses, together with tail QLIKE. For each asset, tail metrics are computed over the top decile of realized-volatility dates in the evaluation sample.

The pattern is consistent across all three ablations. The proposed forecast achieves the lowest overall, high-regime, and tail underprediction losses, as well as the lowest tail QLIKE. The strongest deterioration occurs when the risk-sensitive scoring rule is removed. Taken together, the results suggest that the full design mainly improves stressed-state robustness and protection against severe underforecasting.

\begin{table}[htp]
\centering
\caption{Ablation and robustness checks for stress-aware components. Lower values are better for all metrics.}
\label{tab:focused_ablation}
\scriptsize
\setlength{\tabcolsep}{2pt}
\begin{tabular}{@{}lcccc@{}}
\toprule
Method
& \shortstack[c]{Overall\\ underpred.\ loss}
& \shortstack[c]{High-reg.\\ underpred.\ loss}
& \shortstack[c]{Tail\\ underpred.\ loss}
& \shortstack[c]{Tail\\ QLIKE} \\
\midrule
Proposed forecast         & \textbf{0.0192} & \textbf{0.0474} & \textbf{0.0889} & \textbf{0.6807} \\
No risk-sensitive scoring & 0.0223          & 0.0551          & 0.1000          & 0.7403          \\
No high-state tilt        & 0.0195          & 0.0480          & 0.0900          & 0.6835          \\
No HAR floor              & 0.0195          & 0.0483          & 0.0913          & 0.6887          \\
\bottomrule
\end{tabular}
\end{table}

\subsection{Cross-Asset Comparison}

\begin{figure}[htp]
\centering
\includegraphics[width=0.90\columnwidth,trim=0 8 0 20,clip]
{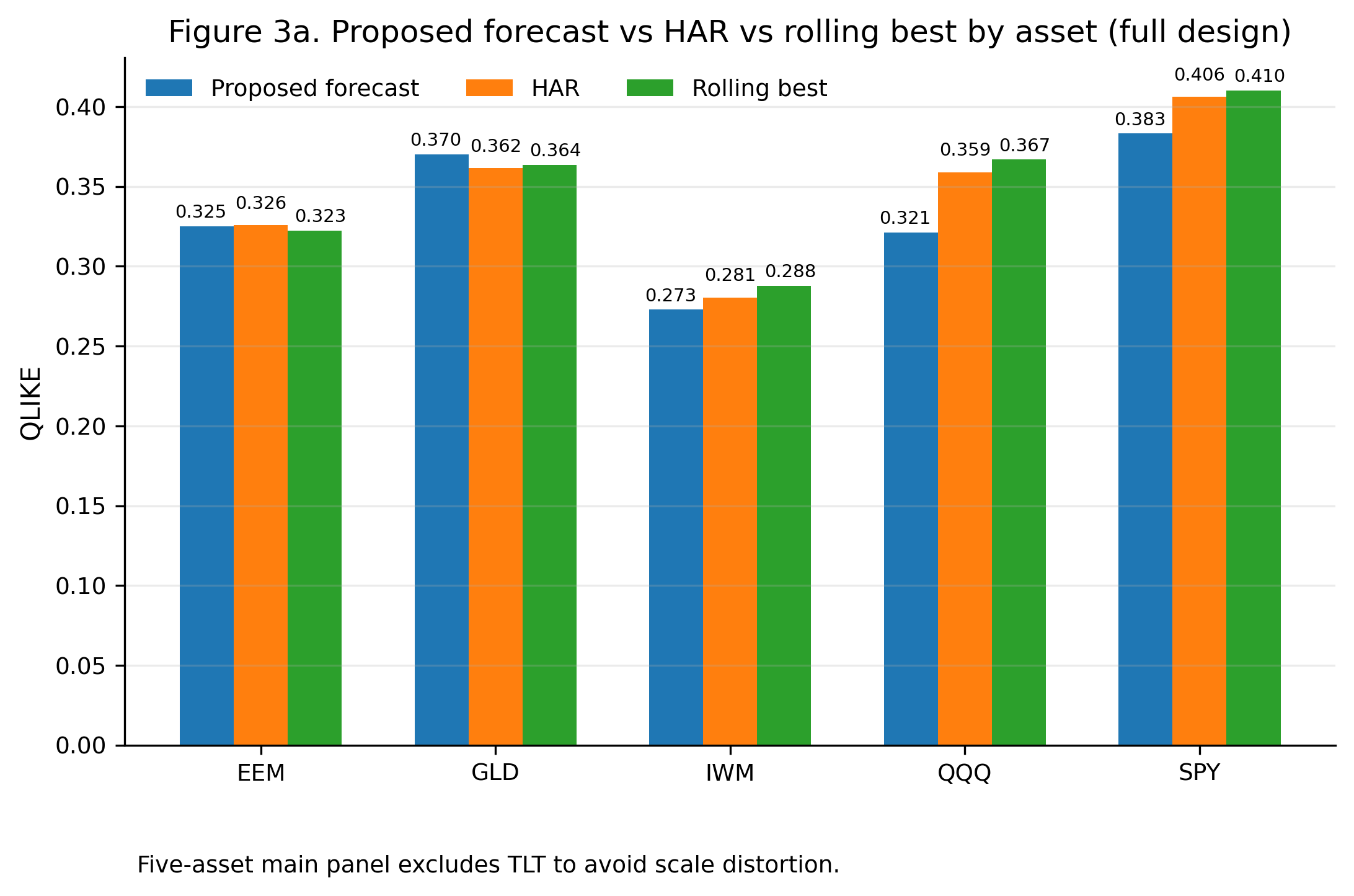}
\caption{Asset-level QLIKE comparison for the five non-TLT ETFs.}
\label{fig:five_asset_compare}
\end{figure}

Fig.~\ref{fig:five_asset_compare} compares the proposed forecast with HAR-RV and the rolling-best benchmark on five non-TLT ETFs. The proposed forecast achieves the lowest QLIKE on IWM, QQQ, and SPY, reducing QLIKE by roughly 5--12\% relative to the rolling-best benchmark on these three assets. On EEM and GLD, the proposed forecast is only slightly above the lower of the two benchmark losses.

\begin{figure}[htp]
\centering
\includegraphics[width=0.90\columnwidth,trim=0 8 0 2,clip]{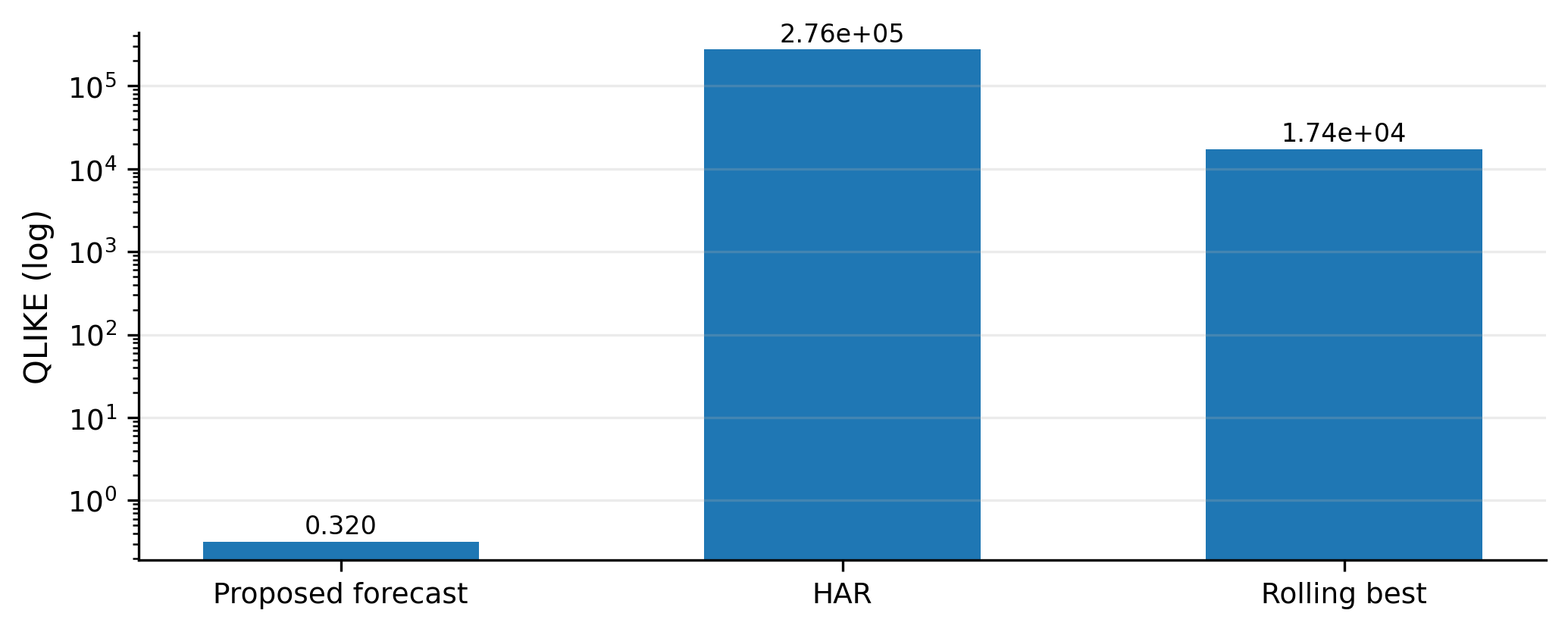}
\caption{TLT robustness comparison on a log scale.}
\label{fig:tlt_robustness}
\end{figure}

TLT is reported separately because baseline failure would distort the main plotting scale. In Fig.~\ref{fig:tlt_robustness}, HAR-RV and rolling-best produce extremely large QLIKE values, whereas the proposed forecast remains near 0.32. Since all methods use the same positive variance floor, this gap reflects repeated severe underforecasting by the baselines rather than a numerical artifact.

\subsection{Relative QLIKE by Regime}

\begin{figure}[htp]
\centering
\includegraphics[width=0.98\columnwidth,trim=0 8 0 20,clip]{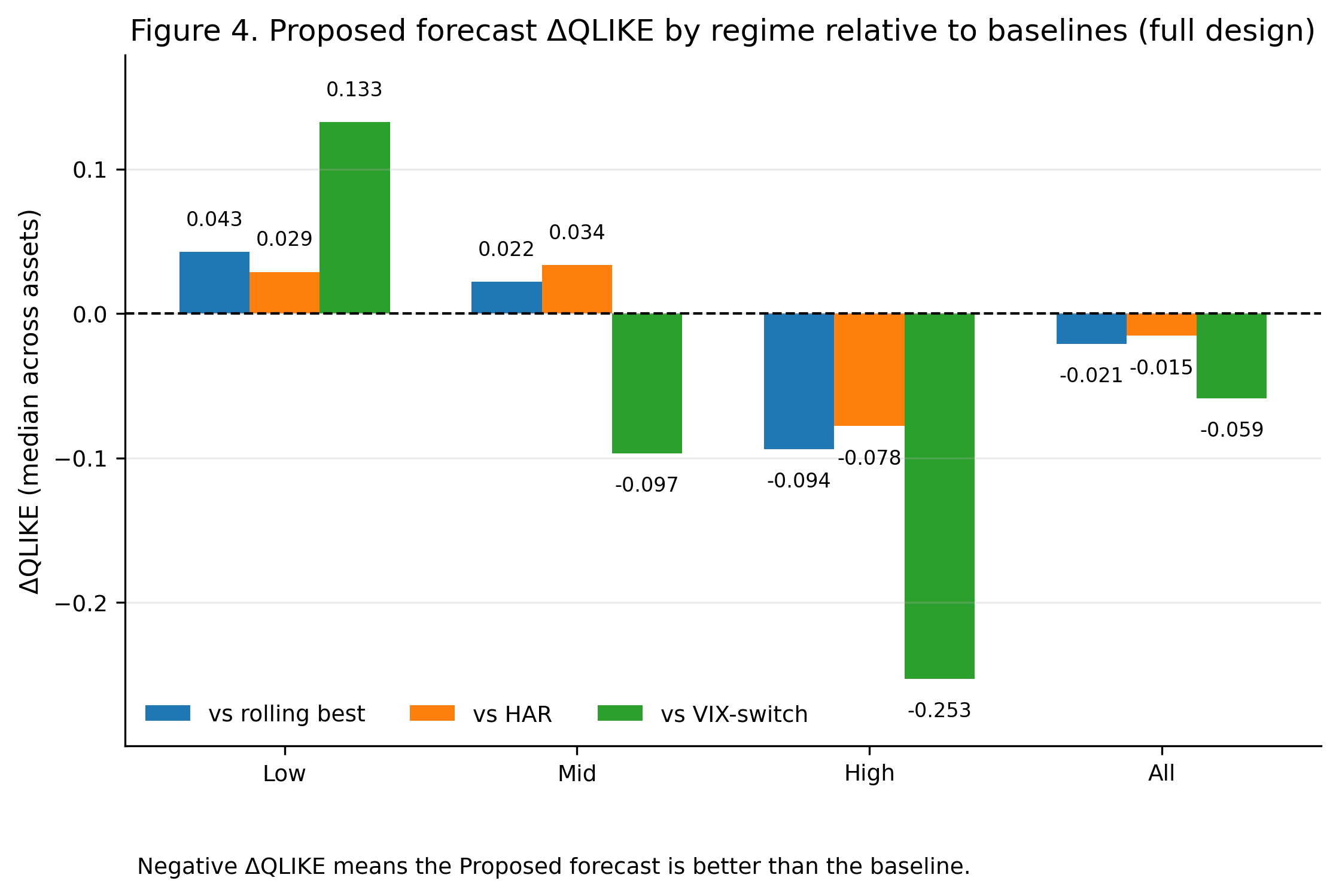}
\caption{Cross-asset median QLIKE differences by regime.}
\label{fig:delta_regime}
\end{figure}

Fig.~\ref{fig:delta_regime} reports the cross-asset median QLIKE difference ($\Delta$QLIKE) between the proposed forecast and three baselines, HAR-RV, the rolling-best benchmark, and the naive VIX-switch rule, where negative values indicate that the proposed forecast performs better. The pattern is asymmetric. Relative to rolling-best and HAR-RV, the proposed forecast is less favorable in the low and mid regimes, but more favorable in the high regime and overall. Against VIX-switch, it underperforms in the low-volatility regime ($+0.133$), but outperforms in the mid, high, and overall samples ($-0.097$, $-0.253$, and $-0.059$).

We also report asset-level Diebold--Mariano tests in Table~\ref{tab:dm_full}. The reported statistics are based on the loss differential defined as proposed minus benchmark, so negative values favor the proposed forecast. Against the rolling-best benchmark, statistical support is selective: the routing forecast is favored on IWM, QQQ, and SPY at the 5\% level. By contrast, the evidence is much stronger against the naive VIX-switch baseline: significant support at the 5\% level appears for five of the six assets, while SPY remains directionally favorable but marginal.

\begin{table}[htp]
\centering
\caption{Asset-level Diebold--Mariano tests.}
\label{tab:dm_full}
\scriptsize
\setlength{\tabcolsep}{4pt}
\begin{tabular*}{\columnwidth}{@{\extracolsep{\fill}}lcccccc@{}}
\toprule
& \multicolumn{2}{c}{Proposed vs Rolling-best}
& \multicolumn{2}{c}{Proposed vs VIX-switch} \\
\cmidrule(lr){2-3}\cmidrule(lr){4-5}
Asset & DM stat & p-value & DM stat & p-value \\
\midrule
EEM & 0.306 & 0.760 & -5.565 & $<0.001$ \\
GLD & 1.353 & 0.176 & -6.097 & $<0.001$ \\
IWM & -2.245 & 0.025 & -3.211 & 0.001 \\
QQQ & -1.977 & 0.048 & -3.611 & $<0.001$ \\
SPY & -2.233 & 0.026 & -1.676 & 0.094 \\
TLT & -1.004 & 0.316 & -4.833 & $<0.001$ \\
\bottomrule
\end{tabular*}
\end{table}

\section{Conclusion and Future Work}

This paper studied next-day ETF volatility forecasting under changing market conditions and proposed a specialist routing framework based on online risk-sensitive scoring and state-dependent forecast combination. The results show that model performance varies substantially across regimes and that the proposed framework improves stressed-regime robustness while remaining competitive overall. Practically, these findings suggest that adaptive routing can outperform reliance on a single fixed model when market conditions shift.

The risk-sensitive criterion is most appropriate when volatility underprediction is costlier than overprediction; it is not a trading objective and may be less suitable under symmetric forecast-error costs. Future work can test alternative volatility proxies, regime definitions, assets, and data-driven state representations.

\smallskip
\noindent\textbf{Code availability:} Code and full hyperparameters are available at \href{https://github.com/TenghanZhong/regime-aware-volatility-routing}{github.com/TenghanZhong/regime-aware-volatility-routing}.

\bibliographystyle{IEEEtran}
\bibliography{references}

\end{document}